# $C_{60}$-BASED COMPOSITES IN VIEW OF TOPOCHEMICAL REACTIONS. II. $C_{60}$ + CARBON NANOBUDS

**E.F.Sheka and L.Kh.Shaymardanova**

Peoples' Friendship University of Russia
sheka@icp.ac.ru

**ABSTRACT**: The current paper presents the second part of the study devoted to composites formed by fullerene $C_{60}$ and single-walled carbon nanotube ($[C_{60}$-(4,4)] carbon nanobud). The formation of composites is considered from the basic points related to the atomic chemical reactivity of the fullerene molecule and carbon nanotube. The barrier that governs the composites formation is determined in terms of the coupling energy $E_{cpl}^{tot}$ and is expanded over two contributions that present the total energy of deformation of the composites' components $E_{def}^{tot}$ and the energy of covalent coupling $E_{cov}^{tot}$. The computations were performed by using the AM1 semiempirical version of unrestricted broken symmetry Hartree-Fock approach.

## 1. Introduction

There have been known a few attempts to synthesize $C_{60}$+CNT complexes that present a single structure in which the fullerenes are covalently bonded to the outer surface of the tubes. For the first time, the covalently bound $C_{60}$ with single-wall CNT (SWCNT) was obtained by means of solid phase mechanochemical reactions [1]. The next time the $C_{60}$+SWCNT complex was synthesized via a microwave induced functionalization approach [2]. It has been used as a component of the photoactive layer in a bulk heterojunction photovoltaic cell. The results indicate that $C_{60}$ decorated SWCNTs are promising additives for performance enhancement of polymer photovoltaic cells. Simultaneously a large investigation has been performed to produce fullerene-functionalized SWNTs, which were termed nanobuds (carbon nanobuds (CNBs) below) and which were selectively synthesized in two different one-step continuous methods, during which fullerenes were formed on iron-catalyst particles together with SWNTs during CO disproportionation [3-5]. It was suggested that the field-emission characteristics of CNBs might possess advantageous properties compared with SWCNT or fullerenes alone, or in their nonbonded configurations.

Computational consideration of CNBs has been restricted so far to two publications [3, 6]. The computations were performed in the framework of the density functional theory (DFT) by using periodic boundary conditions (PBCs) in the restricted closed-shell approximation. A few compositions of intermolecular C-C bonds that form the contact zone on sidewall of the tubes have been considered among which [2+2] cycloaddition turned out to be the most efficient. Moving out of PBCs, considering the formation of CNB occurred in due course of a DA reaction in terms of general energetic scheme presented in Fig.1 of Part 1 [7], and applying broken symmetry approach, which is based on unrestricted open shell approximation and that is more suitable for partially radicalized both fullerene $C_{60}$ and CNTs [8-10], we present another view on the CNBs behavior.

## 2. Computational synthesis of carbon nanobuds

Partial radicalization of CNTs is connected with effective unpairing of odd electrons due to elongation of C-C bonds in comparison with a critical value for their length of 1.395Å[37] similarly to the situation with fullerenes discussed in [10]. Distributed over the tube atoms by partial number of effectively unpaired electrons $N_{DA}$, the electron highlights the map of chemical activity of the tube in terms of atomic chemical susceptibility (ACS) $N_{DA}$. Figure 1 presents the ACS distribution over atoms of two (4,4) SWCNTs shown in Fig.2. The tubes differ by the end atoms that are non-terminated or empty in Fig.2a (tube 1) and terminated by hydrogen atoms in Fig.2b (tube 2).

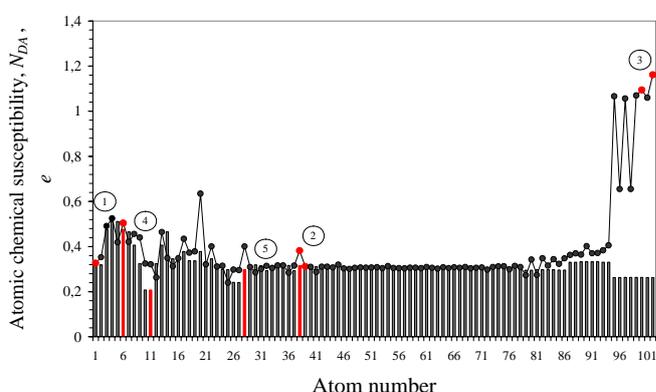

**Figure 1.** Distribution of the **a**tomic chemical susceptibility $N_{DA}$ over atoms of fragments of (4,4) single walled carbon nanotubes with empty (curve with dots) and hydrogen-terminated (histogram) ends [9]. The atom numbering of plottings corresponds to that one over the tubes from the tube caps to their ends. Circled numbers denote atom pairs subjected to further $C_{60}$ addition (see text).

As seen in Fig.1, there are three zones in the ACS distribution related to the tube cap, sidewall, and end, respectively. Accordingly, we have chosen three pairs of target atoms on tube 1 (1, 2, and 3), which are shown by red dots on the curve, and two pairs on tube 2 (4 and 5) shown by red bars. As was stated in [10], the most active atoms of fullerene $C_{60}$ form particularly oriented two hexagons, each atom of which is the target that first meets any addend. The two features related to chemical portraits of CNTs and fullerene molecule make it possible to construct five starting configurations of possible [$C_{60}$+(4,4)] CNBs that are presented in Fig. 3 alongside with equilibrium structures. The starting intermolecular C-C distances were taken 1.7Å. Figures in circles number the corresponding CNBs.

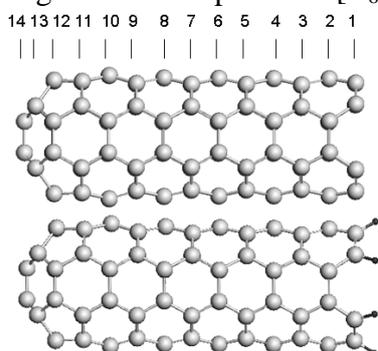

**Figure 2**. Equilibrium structures of (4,4) single-walled carbon nanotube fragments with empty (a) and hydrogen-terminated (b) ends.

Figures 4a and b present changing in the ACS maps of both (4,4) SWCNT and fullerene $C_{60}$ related to CNB 5. As seen in the figure, the attachment of the fullerene molecule to the sidewall of the tube causes only local changing that concerns atoms participating in the formation of the [2+2] cycloaddition. The other part of the atomic activity distribution of the tube retains non-perturbed. This finding evidently favors a multiple attachment of fullerenes to the tube in a superposition manner. Oppositely, the fullerene ACS map changes considerably indicating a significant redistribution of the atomic chemical activity over the molecule atoms after attachment. Red dots on plotted curve in Fig.4b highlight new the most active atoms prepared for the next reaction events.

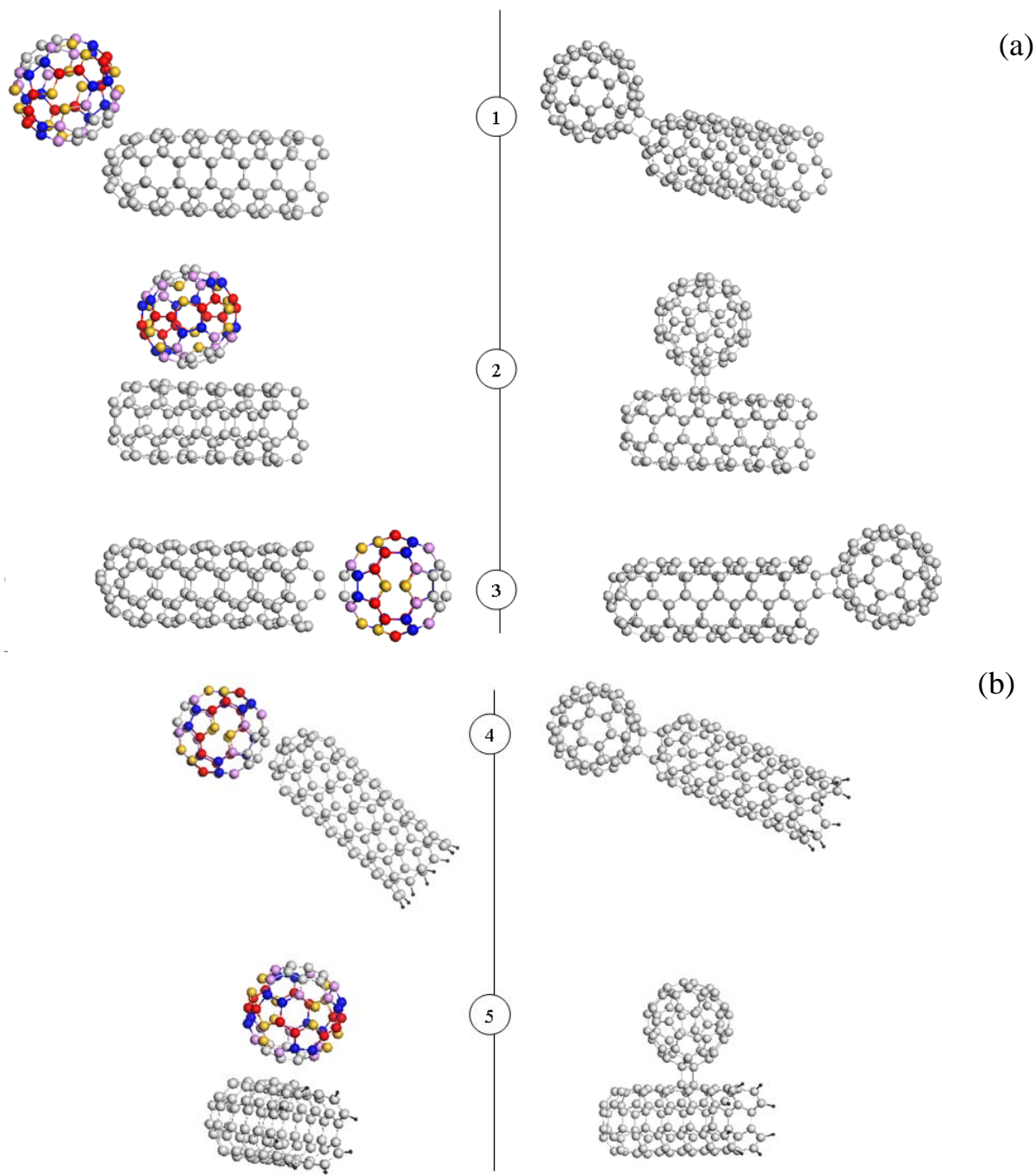

**Figure 3.** Start (left) and equilibrium (right) structures of carbon nanobuds formed by attaching $C_{60}$ to (4,4) single walled carbon nanotube with empty (a) and hydrogen-terminated (b) ends. Circled figures number CNBs (see description in the text).

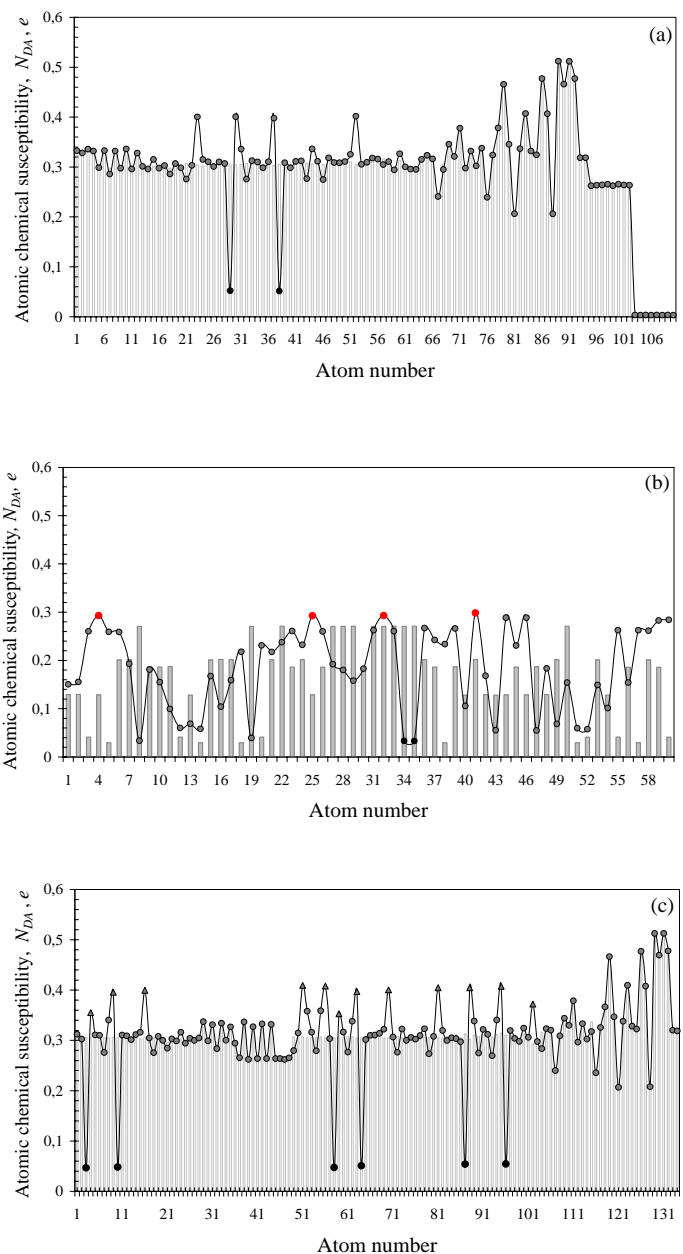

**Figure 4**. Atomic chemical susceptibility evolution of carbon nanobud coponents. *a*. Tube 2 (CNB 5); *b*. Fullerene $C_{60}$ (CNB 5); *c*. Tube 3 (CNB 7) (see text). Histograms present data for the pristine species. Curve with dots plot data related to the formed CNBs.

To check a high tolerance of the tube body to a multiple attachment of the fullerene molecules, two and three $C_{60}$ were attached to an elongated (4,4) SWCNT (tube 3) forming CNB 6 and CNB 7 (see Fig.5). As shown in [9], the tube elongation causes the elongation of the sidewall zone in the ACS map only and does not touch either cap or end atoms regions. That is why the conditions for the formation of CNBs 5, 6, and 7 are identical. All $C_{60}$ molecules are joined with the tube body via [2+2] cycloadditions. Changing in the ACS distribution related to tube 3 is shown in Fig.4c. A clearly seen, the superposition of the three attachments is perfectly exhibited by the map indicating that practically countless number of fullerene molecules can be attached to SWCNTs long enough.

Figure 6 presents a collection of typical [$C_{60}$+(4,4)] CNBs. The set involves multi-attached CNBs by using a multiple attachment of $C_{60}$ molecules to the tube body. Beside this kind of further modification, there are favorable reasons for continuing the modification that involves changing in the fullerene addends. According to the ACS maps of the latter, each attached molecule can be characterized by two types of the most active high-rank-$N_{DA}$ atoms (similarly to the situation with ($C_{60}$)$_2$ dimer [7]) that are marked by red balls in the figure. The first pairs combine the most reactive atoms adjacent to the cycloaddition (below, contact-adjacent or *ca* atoms). Next by reactivity atom's pairs are located in equatorial planes of the molecules (below, equatorial or *eq* atoms). Any further chemical modification of the CNBs via fullerene will depend on the addend size and will be favorable by targeting *ca* atoms by small addends while becoming preferable when targeting *eq* atoms by bulky addends. A predetermined position of *ca* and *eq* atoms makes the modification of CNBs in all cases predictable and controlled.

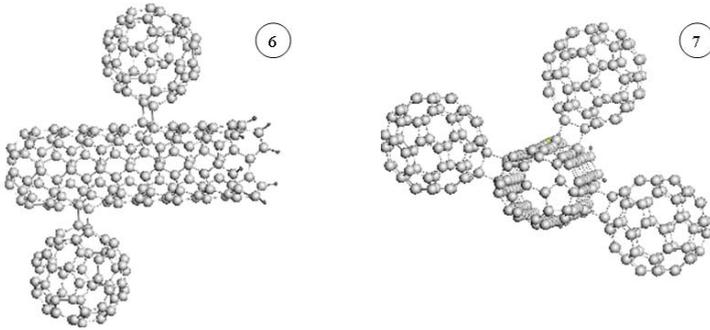

**Figure 5**. Equilibrium structures of [$C_{60}$+(4,4)] carbon nanobuds related to the double (6) and triple (7) attachments of $C_{60}$ to the sidewall of (4,4) single walled carbon nanotube with hydrogen-terminated ends. Circled figures number CNBs.

A detail analysis of the CNB collection presented in Fig. 6 shows that the intermolecular junctions as [2+2] cycloadditions are formed only in the case when fullerene is covalently coupled with the tubes bodies. The junctions in CNBs 1 and 3 are not typical for the cycloaddition in spite of four-atom membership.

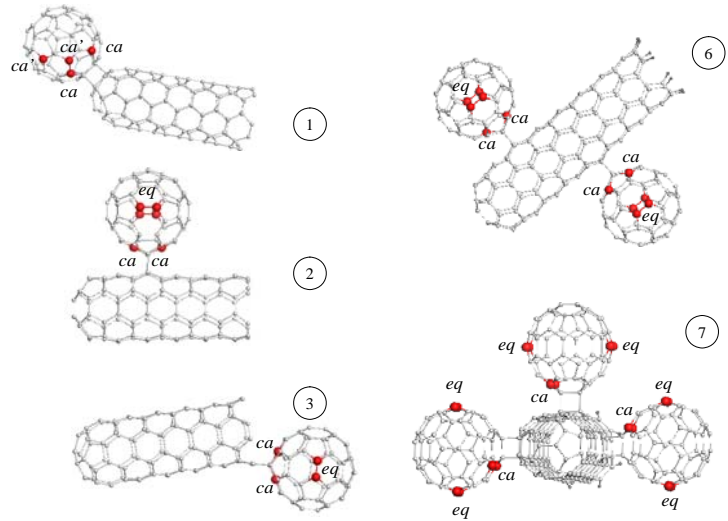

**Figure 6**. Equilibrium structures of [$C_{60}$+(4,4)] Carbon nanobuds. Red balls indicate $C_{60}$ atoms with the high-rank $N_{DA}$ values (see text).

## 3. Energetic parameters and reaction barrier of carbon nanobuds

Energetic characteristics related to the obtained [$C_{60}$+(4,4)] CNBs are presented in Table 1. The coupling energy $E_{cpl}^{tot}$ is determined as

$$E_{cpl}^{tot} = \Delta H_{CNB} - \Delta H_{CNT} - \Delta H_{C_{60}}. \quad (1)$$

Here $\Delta H_{CNB}$, $\Delta H_{CNT}$, and $\Delta H_{C_{60}}$ present heats of formation of the equilibrium structures of CNB, (4,4) SWCNT, and $C_{60}$, respectively.

Similarly to the discussed in [7], we suggest that the total coupling energy reflects two processes that accompany the fullerene attachment to the tube, namely, the deformation of

both CNB components and their covalent coupling. The energy caused by deformation can be determined as

$$E_{def}^{tot} = E_{defCNT} + E_{defC_{60}}, \qquad (2)$$

where

$$E_{defCNT} = \Delta H_{CNT}^{CNB} - \Delta H_{CNT} \quad \text{and} \quad E_{defC_{60}} = \Delta H_{C_{60}}^{CNB} - \Delta H_{C_{60}}. \qquad (3)$$

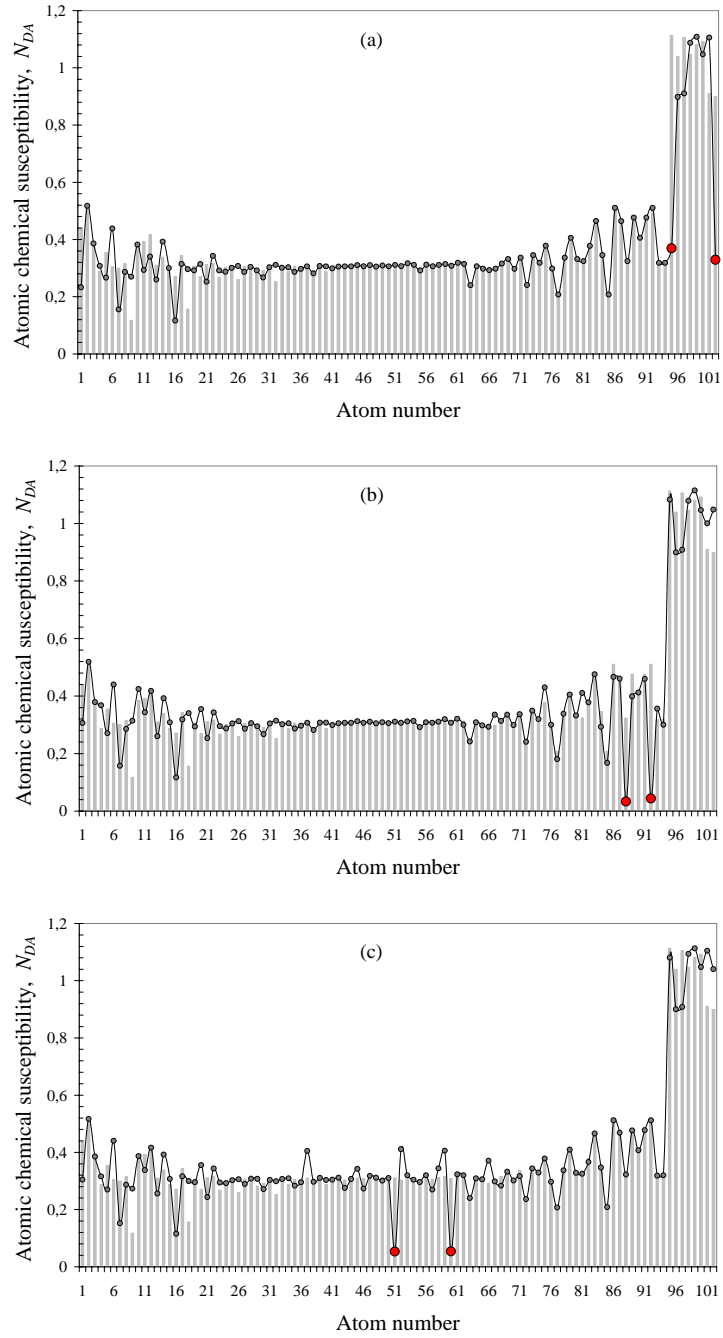

**Figure 7**. Atomic chemical susceptibility evolution of tube 1 when $C_{60}$ is attached to the tube empty end (a), cap (b), and sidewall (c). Histogram presents data for pristine tube. Curve with dots plot data related to the formed carbon nanobuds 1, 2, and 3, respectively. Red dots mark atoms, to which fullerene is attached. Atom numbering corresponds to that one in the output files.

**Table 1.** Energetic characteristics of equilibrium [$C_{60}$+(4,4)] CNT nanobuds, *kcal/mol*

| Nanobuds[1] | $E_{cpl}^{tot}$ | $E_{def}^{tot}$ | $E_{defCNT}$ | $E_{defC_{60}}$ | $E_{cov}^{tot}$ |
|---|---|---|---|---|---|
| 1 (cap) | -36,33 | 51,16 | 10,62 | 40,53 | -87,48 |
| 2 (wall) | -3,38 | 59,64 | 24,64 | 35 | -63,02 |
| 3 (end) | -86,65 | 47,65 | 8,25 | 39,4 | -134,31 |
| 4 (cap) | 3,09 | 114,38 | 62,76 | 51,62 | -111,29 |
| 5 (wall) | -4,26 | 74,33 | 39,26 | 35,07 | -78,59 |
| 6 (wall) | -8,21 (-4,10)[2] | 155,62 | 85,33 (42,66)[2] | 70,29 (35,15)[2] | -163,83 (-81,92)[2] |
| 7 (wall) | -11,02 (-3,67)[2] | 221,64 | 116,59 (38,86)[2] | 105,05 (35,02)[2] | -232,66 (-79,55)[2] |

[1] Figures number CNBs as in Figs. 3 and 5
[2] Data per one attached $C_{60}$ molecule.

Here $\Delta H_{CNT}^{CNB}$ and $\Delta H_{C_{60}}^{CNB}$ present heats of formation of one-point-geometry configurations of the SWCNT and fullerene components of the equilibrium configurations of the studied CNBs. Accordingly, the chemical contribution into the coupling energy is determined as

$$E_{cov}^{tot} = E_{cpl}^{tot} - E_{def}^{tot}. \tag{7}$$

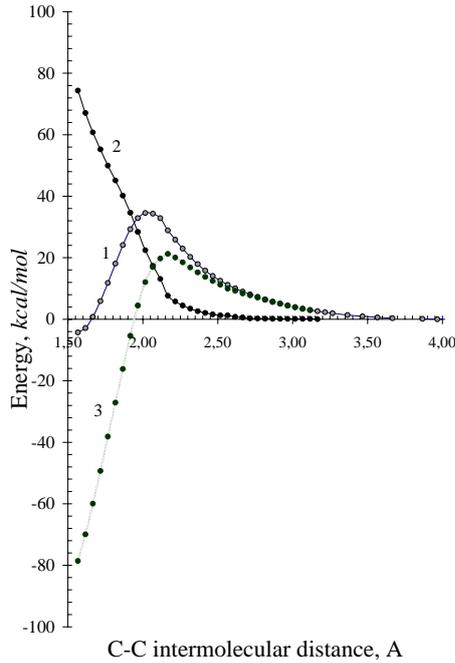

**Figure 8.** Profile of the barrier of the [$C_{60}$+(4,4)] carbon nanobud decomposition. 1. $E_{cpl}^{tot}$; 2. $E_{def}^{tot}$; 3. $E_{cov}^{tot}$.

Analyzing data in Table 1, two distinct features should be emphasized. The first concerns the difference in the behavior of CNBs, fullerene component of which is attached to either the cap or end atoms of the tube. The second is related to a close similarity in the properties of CNBs with fullerene attached to the tube sidewall. As seen in the table, the empty end of the tube, once the most active according to the ACS map in Fig. 1, provide the fullerene attachment with the biggest coupling energy and the smallest deformation energy related to the tube. It means that at any contact of such tube with fullerene (as well as with any other addend), the first attachment occurs on the tube open end. The next events will take place at the tube cap. The coupling energy decreases by 2.4 times while the deformation energy slightly increases. After these two events, there comes a turn of the tube sidewall, but the coupling energy decreases by ~26 times when the deformation energy increases three times. The three events are quite superpositional as can be seen in Fig. 7. Each addition concerns a strictly local area; so that highly active attachments to the end and cap region should not prevent from covering the main tube body by multiply attached fullerenes.

Obviously, the end- and cap attachments do not contribute much to the massive deposition of $C_{60}$ on CNTs, the main part of which is formed by fullerene molecules attached to the tube extended sidewall. The configuration of CNB 5 with a single attachment of fullerene has been chosen for determining the reaction barrier for this most typical case. Figure 8 presents the dependence of the basic energetic characteristics of the CNBs according to expressions 1, 2, and 4 related to the current C-C intermolecular distance $R_{CC}$. As in the case of fullerene dimers discussed in [7], the barrier energy computation has been started from the equilibrium configuration of CNB 5 followed by a stepwise elongation of two C-C bonds that provide intermolecular contact via [2+2] cycloaddition. A deep parallelism in the behavior of fullerene molecules either singly bound to the tube body or alongside with other molecules in the case of multiple attachments provides a good reason to expect the same parallelism between the energy dependences of these molecules as well.

Three plottings in Fig.8 behave quite similarly to that ones of $C_{60}$ dimer [7], which could be expected, since the contact zones in both cases are presented by similar [2+2] cycloadditions. The maximum of the coupling energy $E_{cpl}^{tot}$ is located at 2.05Å so that at starting distances exceeding this value CNT and $C_{60}$ form a charge transfer complex, where CNT donates and $C_{60}$ accepts electron under photoexcitation, with equilibrium spacing between partners of 4.47Å and coupling energy of -0.06 kcal/mol. The coupling energy $E_{cpl}^{tot}$ in Fig.8 can evidently be divided into $E_{def}^{tot}$ and $E_{cov}^{tot}$ components of the same type as those related to $C_{60}$ dimer. However, the difference in numerical values of the two components at starting point results in much shallower minimum of $E_{cpl}^{tot}$ in the case of CNB and, thus, in lesser barrier for the CNB decomposition. The feature has caused a suspicion that the difference in the behavior of $(C_{60})_2$ dimer and $[C_{60}+(4, 4)]$ CNB is connected with a particular topological character of joining $C_{60}$ not only to the same molecule but to SWNT as well. If so, a similar event should occur when $C_{60}$ is accommodated in the close vicinity of a planar graphene sheet, once different from both $(C_{60})_2$ dimer and $[C_{60}+(4, 4)]$ CNB. The computations performed in the study have proven this prediction quite convincingly [11].